\documentclass[a4paper,12pt,leqno]{article}
\usepackage{amssymb,amsmath,amsfonts,amsthm}
\usepackage[a4paper,margin=2cm]{geometry}

\usepackage{newcent}
\usepackage[scaled=1.1]{newtxmath}

\usepackage{graphicx}
\graphicspath{ {./images/} }
\usepackage[usenames,dvipsnames]{xcolor}
\usepackage{longtable,booktabs,array,calc}
\usepackage{enumitem}
\setlist{itemsep=1pt,topsep=3pt}
\usepackage{titlesec}
\titleformat{\paragraph}[runin]{\normalfont\normalsize\bfseries}{}{0pt}{}
\titlespacing*{\paragraph}{0pt}{1.5ex plus 0.5ex}{1em}

\usepackage{fontawesome}

\providecommand{\tightlist}{%
  \setlength{\itemsep}{0pt}\setlength{\parskip}{0pt}}
\usepackage[round,authoryear]{natbib}
\usepackage[pdfencoding=auto]{hyperref}

\usepackage{tcolorbox}
\newtcolorbox{aiprompt}{
  colback=gray!5,
  colframe=gray!40,
  title=\textbf{Prompt},
  fonttitle=\bfseries,
  boxrule=0.5pt,
  left=5pt,
  right=5pt,
  top=3pt,
  bottom=3pt
}

\theoremstyle{definition}
\newtheorem{definition}{Definition}
\newtheorem{example}{Example}

\theoremstyle{plain}
\newtheorem{proposition}{Proposition}

\newcommand{\DLI}{\mathrm{DLI}}
\newcommand{\DE}{\mathrm{DE}}
\newcommand{\Norm}{\mathrm{Norm}}
\newcommand{\Ax}{\mathrm{Ax}}
\newcommand{\Dir}{\mathrm{Dir}}
\newcommand{\voc}{\mathrm{voc}}
\newcommand{\Proof}{\mathbf{P}}
\newcommand{\Min}{\mathrm{Min}}

\title{When Proofs Are Abundant, What Is a Paper For?\\[.3em]\Large Intension and Scientific Chains of Thought\\ in the Age of AI Mathematics}
\author{Charles-Albert Lehalle\thanks{CMAP, Ecole Polytechnique, IP-Paris -- {\tt charles-albert.lehalle@polytechnique.edu}}}
\date{Nov 4, 2026 -- Work in Progress (v2.3.1)}

\begin{document}
\maketitle

\begin{abstract}
  AI systems now produce formally verified proofs of celebrated open problems, and the cost of obtaining a correct proof is falling toward zero. If correctness becomes cheap, what is left for a research paper to communicate? I argue that a paper is not exhausted by a verified statement. It is a narrative exposed to the criticism of a community, made of an intension (the author’s choice of problem and perspective), a scientific chain of thought (the reasoning steps leading to the result), and the arguments supporting each step. Most published mathematics is judged on these two elements rather than on correctness alone, and machine-generated formal scripts convey neither. Borrowing from logic and type theory, I formalise the first as a Derivation-Level Intension and the second as a Directional Extension: a decomposition of a proof into cut-points whose axiomatic footprints reveal its weak links, where the result could be strengthened, and its strong links, which can be transported to other theories.
  
  Moreover, I examine five roles AI can play in the research process, including bibliographical search and critical judgment, and show that none of them removes the need to declare these two elements. I mention ways these two concepts could be useful for Journals and for training of young mathematicians in a AI era.
  If proofs become abundant, the scarce resource is not correctness but a community’s shared understanding of why a result matters and how it could be pushed further. 
\end{abstract}
\tableofcontents

\section{The need for humans doing mathematics}


Media outlets have lately fixated on the dramatic triumphs of artificial intelligence systems as they claim to gave dismantled long-standing mathematical problems \citep{openai2026navierstokes}.
The immediate administrative fallout, however, has been less triumphant: scholarly journals have been so thoroughly overwhelmed by automated submissions that several have taken the drastic step of suspending new manuscript evaluation entirely \citep{esaimps2026announcement}.

This institutional crisis forces us to confront once again the epistemic function of the scientific paper. Though widely treated as a modern pathology peculiar to the machine age, the question itself is part of a longer historical continuity. As André Joyal remarked at the outset of a 2018 address \citep{joyal2018hott}:
``\emph{[Voevodsky aimed to] develop proof assistants based on univalent theory.
  This was a very important aspect of [his] program because he wanted to give people some kind of concrete tools for doing mathematics. And that could revolutionize in some way the way we do mathematics, because we would be able to write proofs on a computer, and possibly write papers in the same way.
  If the proof would be automatically checked or verified by the univalent system, by the proof assistant, therefore a referee will not need to check if the result is proved or not.
  I mean, you would submit a paper with the proof and the verification, and the referee... the job of the referee will not be to check if the proof is correct [...] so his job would be to decide if the result is interesting or not!
  Which is, in some sense, a better kind of thing to do in many ways. That could revolutionize reading mathematics, because by speeding up the publication process in many ways.}''

\medskip

One might belive that journals and research seminars exist primarily as proof checkers; gatekeepers whose sole mandate is to stamp a submission as either valid or invalid.
Such statements miss the true nature of scientific progress. Publications and talks are not mechanical filters; they are the forums through which a research community exchanges ideas, offers criticism, and draws lessons from new developments.

Today, Yehuda Rav's imaginary Pythiagora oracle, a desktop companion capable of delivering an instant verdict on any proposed theoreme \citep{rav1999we}, has taken shape in agentic systems combining large language models with automated code checkers. Yet, even in this era of mechanized verification, the act of sharing proofs and results remains as vital as ever since the role of the research paper is to serves as an impetus for broader intellectual growth. When authors present a result to their peers, the dialogue flows in two directions:
\begin{itemize}
\item \emph{Justifying the Problem's Value}: Authors do not simply ask for a validation of their logical steps.
  Above all, they contend that the problem itself is worth solving, highlighting its practical applications or the illuminating connections it establishes between core concepts.
  A reviewer’s primary task is to grasp this broader context and evaluate the result's true merit for the community.
\item \emph{Exposing the Machinery of Reasoning}: Authors unpack their internal methodology, presenting their argument as a sequence of distinct conceptual building blocks.
  They intentionally highlight vulnerable steps, inviting the community to strengthen them and render the result more powerful, while showcasing exceptionally strong techniques designed to be adapted for different questions.
\end{itemize}
This ongoing debate over an author's angle and key conceptual moves is what drives the collective forward, generating far richer ideas than any isolated statement of truth ever could.

Some may fall into the trap of believing that with Pythiagora at our fingertips, human intuition is no longer needed to formulate new questions. It is easy to imagine an autonomous machine inventing important scientific problems on its own. But this is an illusion. Discovering meaningful new problems requires for a system an awareness of what lies ``outside'' the horizon of what has already been modelled.
Neverthless to be reachable by the system they need to be ``inside'' the formalized concepts, hence if the mechanism of answering to a question by true or flase is free, they are already solved.
%
The situation brings to mind a classic economic paradox, analogous to the Grossman-Stiglitz result on informationally efficient markets \citep{grossman1980impossibility}: if the cost of generating formal proofs drops to zero, the incentive to prove vanishes, unless human agents intervene to select which problems are worth formulating in the first place.

A tempting resolution is to imagine a new division of labor where engineers serve as the human operators of Pythiagora. In this scheme, engineers remain safely outside the formal machine, picking fruitful starting points in Bouveresse's vast Land of Possibles \citep{bouveresse1988pays}, while leaving the mechanical derivations to the automated oracle.

Yet this separation is an illusion. An operator cannot simply feed raw reality into a formal machine. Engineers must continuously model the domain, formalize physical sensors so they speak the system’s language, and construct ideal descriptions of actuators within the same formal framework. Because no model can ever capture reality with absolute precision, the process becomes an iterative, back-and-forth dialogue, placing Pythiagora directly inside the design loop.

Naturally, as these technical demands grow, the operators will specialize. They will form distinct sub-communities, delegating aspects of their modeling tasks to colleagues who focus either on reasoning with formal components or on exploring alternative formalisms.

But we need not speculate about the future of this workforce: we already know who these specialists are. The former are what we call applied mathematicians; the latter, pure mathematicians.

\section{Proofs evolved through ages, and why do we publish}

Before going further in the discussion towards formal definitions, it is worthwhile to investigate of the potential shift of paradigm these new tools are pushing us to make, if any.
Let me start by an historical perspective.

\subsection{A biased sketch of the evolutions of proofs}

Mathematics are the language of science; their rigor, even if in some cases tainted with undecidability,\footnote{It is established since the 1950ties that some mathematical theories contain non decidable propositions \citep{Tarski1953Undecidable}.} is such that other sciences consider that the mathematical implications of a well-written statement are true when the premises of the statement are true.\footnote{This argument is compatible with the view developed in \cite{lyons2026mathematics}: ``\emph{This brings us back to the purpose of mathematics. Its importance lies in the relationship between mathematical understanding and the world: in recognising which structures matter, developing ways to reason about them, and connecting the resulting knowledge to scientific and human needs.}''}
The concept of ``mathematical proof'' has nevertheless evolved though centuries, and more recently, decades.
A example has been \emph{The Nine Chapters on the Mathematical Art} (Jiǔzhāng Suànshù), \citep{Shen1999} compiled between the 2nd century BCE and 1st century CE,\footnote{This corpus as been considered as ``incorrect mathematics'' for centuries until, between 1960 and 1980, investigations led to the conclusion that this collection of recipies where indeed very advanced \emph{algorithms} that we would call today constructive proofs. 
The translators in French of The Nine Chapters \citep{ChemlaShuchun2004} write in the preface: ``We can therefore say that the classical mathematics of ancient China and Euclid's Elements represent two radically different styles of mathematical reasoning. The comparison gives us a crucial lesson here: there is no single path of development that mathematicians should have followed, wherever they might be. What makes Euclidean axiomatic systems strong is that they identify what must be assumed so that different conclusions can be deduced. The resulting weakness stems from the difficulty of finding axioms that satisfy everything at once.''.
} 
and a more recent example is the debate around the formalization of the probability theory that ended in 1933 only (\citep{Kolmogorov1956} for the English version).

A very important step in the recent standardization of mathematical proofs started with ``Hilbert's Program'' in the 1920ties \citep{Zach2020}.
The discovery of undecidable statements in some mathematical theory gave birth to constructive mathematics, that led to the evolution of logic on Type theory and Category theory, and more importantly for the current debate to formal theorem proving, from the 60ties \citep{deBruijn1970} to recent years through different tools (Isabelle, Coq, etc.).
A notable tool of this family is Lean \citep{deMoura2015}; that became later commonly used by probabilistic exploration systems embedding Large Language Models \citep{Bansal2019,Han2021}.\footnote{In essence, the goal of such machineries is to explore all the possible paths the expressivity of Lean offers with some probabilistic guidance to direct the search towards directions that have not too low probability of being exact. See \citep{Jiang2026} for a recent survey.}

\paragraph{The ``Navier-Stokes moment''.}
These systems have recently obtained unexpected success with a famous Millennium Problem: Navier–Stokes existence and smoothness \citep{openai2026navierstokes}, and the ``dying percolation conjecture'' \citep{PostCutoff2026}.

The standard criticisms on ``AI-generated'' Lean proofs \citep{WeatherallWolfson2026}, summarized in the title of a blog post by: ``\emph{AI solved one of math's hardest problems. Humanity learned nothing (so far)}'' \citep{Brumfiel2026}, can be read as
\begin{enumerate}
\item Lean scripts do not convey any purpose, 
\item they mix the chain of thought and the arguments to support components of this chain so that there is no human way to judge what is what, so far.\footnote{It is possible that in the future, Lean-reading assistant will help human to extract some meaningful chains of thought from this long script of Lean script.} 
\end{enumerate}

\subsection{Why do we publish?}\label{sec:history}

Coming back to publications: most research papers in mathematics are not solving a true or false theorem statement that has been elected by the community of mathematicians as an important one.\footnote{Besides, a problem like Navier–Stokes has been elected because the community thought that on the way to prove it, the community will learn enough to go further, discover new important area of research, whereas it is exactly what a Lean script does not.}
Most papers are attempts to shed light on an area of mathematics that, from the viewpoint of the authors, is worthwhile to formalize and solve. 
They defend it by disclosing their intension and their chain of thought. Both are defined with accuracy later in the paper. Sometimes, one component of the chain of thought is solved an innovative way that justifies the publication of the whole chain. In such a case, the paper can then be considered as a demonstration of ``this component usefulness in context''.

As strange as it seems, research papers with a vague target are more difficult to emulate with LLM-driven systems, simply because the intension is a crucial element of these papers. 
Nobody needs to convince the community of mathematicians that solving Navier–Stokes or the dying percolation conjecture is important; research starts after it has been declared important. 
For most other papers, however, a substantial portion of the discussion with referees and journal editors centers on the Derivation-Level Intension of the paper, the conceptual angle chosen by the authors, which can be precisely formalized using Category Theory (as defined in Appendix \ref{sec:formal}).%
\footnote{For instance, a large part of my discussion to publish \citep{CL} was about splitting or not the paper in two, because our correspondents mostly thought the intension would be clearer (one paper with the theory and one paper with the application). With my co-author, we defended that they were going together, partly because the application was a way for us to show very clearly why the theory deserved to be published.
Similarly, for \citep{Lachapelle2016}, the major point of discussion was whether your chain of thought should be written in probabilistic terms or in partial differential equation terms.
In both cases: no question on the correctness of the proofs, but on the Derivation-Level Intension or on the chain of thought.}

In \emph{The Science of The Artificial} \citep{Simon1969}, Simon makes the difference between understanding and acting; in our case a paper is a chain of reasoning in action. 
When Anne Alombert says ``{after the automation of practical skills, we are witnessing the automation of thinking}'' \citep{Alombert2026}, I would replace it by ``\emph{after the automation of practical skills, we are witnessing the automation of reasoning}''. Such that, it applies exactly to research papers: the act of reasoning is automated, but the act of understanding is what should continue to be conveyed via a paper and a talk presenting this paper.
\medskip

The material shared by researches with their community targets to
\begin{enumerate}
    \item propose a narrative (i.e. one intension + a scientific chain of thought + arguments supporting part of the chain of thought)
    \item such a way that it is exposed to the criticism of the community.
\end{enumerate}
It is via collective critical thinking, that can be seen as a debate, that the research is validated or not, that it can foster new ideas and directions, and that the intension is pursued further, or mixed with other ideas by the authors or other researchers.
\medskip

\section{Two formal notions: Derivation-Level Intension and Directional Extension}\label{sec:formal}

This section is the core of the paper. It proposes to formalise, with the help of logic, type theory and category theory, what researchers should share in a paper and during talks exposing their work, particularly if AI systems and LLMs have been used during the process of research. In a world of ``abundance of proofs'' \citep{Tao2026Math20}, scientific communities have to decide how to share the results of research, whatever the means used to produce them, in a way that allows a field to continue to solve more problems.
The proposal is to articulate the description of the outcome of a research project around two concepts: the positioning of the work in the field, and the chain of reasoning chosen by the researchers.

\subsection{What a paper conveys}

The starting point of the thesis defended in this section (aligned with \citep{Sanderson2026}) is that a scientific paper is made of
\begin{itemize}
    \item an intension,
    \item a scientific chain of thought,
    \item a sequence of arguments supporting each element of the chain of thought.
\end{itemize}
The arguments supporting the chain of thought are often referring to external resources.
Typical examples are: a theorem which proof is not repeated in the paper, or a straightforward transposition of a relation to a specific metric, or a well known fixed point theorem on a specific space, etc.\footnote{Example of a section of such a chain of thought is: ``\emph{As almost always the case for this kind of results, the argument of proof consists in applying a fixed point argument of Schauder type and requires therefore compactness properties. The
main difficulty is the control of the time regularity of the evolving measure $\mu$. This regularity
is related with the time regularity of the optimal controls. In the first order case, it is not an
issue because the optimal trajectories satisfy a Pontryagin maximum principle and thus are
(at least) uniformly $C^1$. In the second order setting, the Pontryagin maximum
principle is not so simple to manipulate and a similar regularity for the controls would
be much more heavy to express. We use instead the full power of semi-concavity of the
value function combined with compactness arguments (see Lemma 5.4 below). The main
advantage of our approach is its robustness: for instance stability property of the solution is
almost straightforward.}'' from \cite[Proof of Theorem 5.1]{CL}.}
The first two items are the object of the two notions below; the third is the part that can be delegated to external references, to computations, or to machine-checked scripts.

\subsection{The two notions, in words}\label{sec:announce}

\emph{The Derivation-Level Intension is the derivation; the Directional Extension is its publication structure.}

\paragraph{Derivation-Level Intension (DLI, or intension).}
I propose to use \emph{intension} for the choice of the problem by the scientists, which they propose for review to their peers in writing the paper. The term is borrowed from logic, with a legacy going back to Carnap, and has to be borrowed from type theory and category theory to get a grasp on proofs as objects. It covers the chosen angle, and it has to be reflected in the semantics used to expose the problem. Following Joyal, this is what enables the reviewers to judge ``if the paper is interesting'' for their community.
In the language of mathematics, it should rather be named \emph{Derivation-Level Intension}, to reflect that it covers the semantics of arguments and proofs (without $\beta$-reduction). It is the derivation of the shared results, taken up to bureaucratic reordering of independent steps, together with the vocabulary of its intermediate statements. It answers: \emph{from which angle are these results presented?}
An example is the Pythagorean theorem:
authors may choose to establish it via area dissection (a proof $P_1$) or via similar triangles and segment ratios (a proof $P_2$). These represent two distinct derivation-level intensions.
Each choice mobilises a different field or modelling, even if both reach the conclusion that the theorem is true. Making the community aware of this choice is what makes further progress possible; it also informs perspectives in terms of applications.

\paragraph{Directional Extension (DE, or explained scientific chain of thought).}
The intension is not the only aspect of the research requiring discussion with the community to foster progress: an important point is the structure of the reasoning, presented in such a way that it exposes the strengths and weaknesses of the results.
Weaknesses are the links of the chain that prevent the result from being stronger; they tell the community where to put efforts to improve it. Strengths are links of the chain that are uniquely strong, so that in the authors' view they can be reused for other results. Since the qualification of ``weak'' and ``strong'' is subjective, it is fruitful to share it with the community, via the scientific review process and talks at conferences, to get feedback and establish a collective view on these points.
The language of logic and type theory suggests the name \emph{Directional Extension}, a formal description of the proof with cuts, which I use as a synonym for scientific chain of thought. It describes a published proof through a sequence of exposed cut-points, or lemmas, together with their axiomatic footprints and transport classes. It records which intermediate statements of the derivation the author exposes as stand-alone lemmas, which hypotheses each of them uses, and in which other settings each can be transported. It answers: \emph{what can the community extend and reuse, and where?}
This does more than provide a convenient partition of the reasoning: it preserves a ``message'' by the authors, namely \emph{here are the weak links and here are the ones we believe to be reusable}. Category theory is a suitable framework, since it allows cuts in a proof to be preserved, ideally in such a way that segments are as independent as possible in terms of the concepts and notions they use.

\paragraph{How they fit.} The DLI fixes the cut points that are available in the derivation; the DE selects which of them are published as lemmas. Two papers may share a DLI and still differ in DE (Proposition~\ref{prop:de}).

\paragraph{What this paper does not claim.} It does not claim that every pair of proofs differs in DLI, that DLI or DE are the only relevant invariants, or that either predicts a mathematician's choice. It claims that once a background theory and a calculus are fixed, each notion is a definite object, so that a mathematician can say, in checkable terms: \emph{here is my goal in choosing to work on this theorem, and here is the structure of my proof, in such a way that the community can extend it and reuse it}.

\subsection{Setting}

I am not an expert in the fields I mobilise to provide these attempts at definitions. The aim of this section is to propose definitions and to request comments and improvements.
\smallskip

Throughout, I fix a formal language $L$, a set $\Sigma$ of named axioms formulated in $L$ (the background theory), and a proof calculus $K$ (for instance natural deduction) equipped with cut elimination. $\Norm(D)$ denotes the cut-free form of a derivation $D$ obtained by this procedure, which is assumed to have the Church--Rosser property.

\subsection{Derivations and Derivation-Level Intension}

Fix a language $L$, a set of named axioms $\Sigma$ in $L$, and a deduction calculus $K$ satisfying the Church--Rosser and strong normalization properties for cut elimination ($\beta$-reduction). A \emph{derivation} is a directed acyclic graph $D$ of formula occurrences, whose hyperedges (with ordered premise ports) are inference steps of $K$ and whose leaves are named axioms from $\Sigma$ or hypotheses. Natural-deduction derivations are classically trees. 
Allowing a sub-derivation to be shared by several later steps gives the dag-like presentation standard in proof complexity, 
and graph-like presentations of proofs also appear in proof nets, 
combinatorial proofs 
and categorical coherence. 
The graph presentation lets parallel sub-arguments be combined, which both proofs below do.

A \emph{paper derivation} is a derivation with a distinguished set $R$ of nodes, the \emph{shared results} $T_1,\dots,T_k$ (a main theorem and its companions). Its \emph{trimmed form} $D^{\mathrm{trim}}$ is the sub-DAG induced by the nodes of $D$ from which at least one node in $R$ is reachable. Definitions are kept \emph{folded}: a defined symbol is an atom, and its defining axiom is a named leaf; we never unfold ($\delta$-unfold) a definition. For a formula $S$, $\voc(S)$ is the set of non-logical symbols of $L$ (primitive or defined) occurring in $S$.

\begin{definition}[Derivation equivalence $\approx$]\label{def:approx}
$\approx$ is the smallest equivalence relation on derivations with the same hypothesis and conclusion such that
\begin{enumerate}\tightlist
  \item isomorphic DAGs (preserving inference rules and the order of premises at each inference hyperedge) are equivalent, so the chronological ordering of independent sub-derivations is irrelevant;
  \item removing a sub-derivation that does not reach any shared result (\emph{trimming}, replacing $D$ by $D^{\mathrm{trim}}$) preserves the class;
  \item cut with an identity derivation, re-association of cuts, and bureaucratic conjunction packaging (introducing and immediately projecting from a conjunction solely to bundle parallel premises) preserve the class (the administrative laws of composition).
\end{enumerate}
$\approx$ contains no general $\beta$-reduction (cut elimination) and no $\delta$-unfolding. Any choice of $\approx$ meeting the stated minimal requirements (items 1--3, and contained in $\beta$-convertibility) is admissible; the framework is parametric in it.
\end{definition}

\begin{definition}[Derivation-Level Intension]\label{def:dli}
The \emph{Derivation-Level Intension} of a paper derivation $D$ is its class
\[
\DLI(D) \;=\; [D]_{\approx},
\]
together with its derived invariant, the \emph{vocabulary of the angle}
\[
L_D \;=\; \bigcup_{\substack{S\; \text{formula}\\ \text{occurrence in } D^{\mathrm{trim}}}} \voc(S)\ \subseteq L .
\]
Since $\approx$ preserves formula occurrences in the trimmed form up to isomorphism (and bureaucratic conjunction packaging introduces no new non-logical symbols), $L_D$ is a well-defined function of $[D]_\approx$.
\end{definition}

Write $\Norm(D)$ for the $\approx$-class of the cut-free ($\beta$-normal, $\delta$-folded) form of $D$.

\begin{proposition}\label{prop:norm}
(i) $\DLI(D_1)=\DLI(D_2)$ implies $\Norm(D_1)=\Norm(D_2)$. (ii) The converse fails: if $D$ contains a cut on a lemma $S$ whose vocabulary does not occur in the cut-free derivation $D^\circ$, then $\Norm(D)=\Norm(D^\circ)$ while $\voc(S)\subseteq L_D\setminus L_{D^\circ}$, so $\DLI(D)\neq\DLI(D^\circ)$.
\end{proposition}

\begin{proof}
(i) holds because $\approx$ is contained in $\beta$-convertibility, and $\Norm$ is invariant under $\beta$-convertibility by the Church--Rosser property of $K$. (ii) is immediate from the definitions of $L_D$ and of cut elimination.
\end{proof}

\noindent\textbf{Plain-English meaning.}
Proposition~\ref{prop:norm} explains why a proof cannot be reduced to its fully expanded, cut-free form without losing its scientific point. Part~(i) is a sanity check: if two proofs share the same Derivation-Level Intension (differing only in bureaucratic ways, such as the chronological order in which two independent steps are written down), then eliminating all intermediate lemmas reduces them to the exact same elementary derivation. Part~(ii) states the crucial converse failure: when we eliminate cuts ($\beta$-reduction) to normalize a proof $D$ into $D^\circ$, we erase the intermediate lemmas $S$ and the conceptual vocabulary $\voc(S)$ that the author introduced to bridge the problem. Two derivations can therefore collapse to the same low-level normal form while having genuinely different DLIs. Preserving the DLI prior to cut-elimination is what records the researcher's chosen conceptual angle—the perspective that makes the proof intelligible and interesting to the community.
\medskip

Two linear presentations of a proof of $A\wedge B$ that only exchange the chronological order in which the independent sub-derivations of $A$ and $B$ are carried out yield isomorphic DAGs and have the same DLI, even though their linear proof scripts differ.

\subsection{From informal claims to formal objects: splittings, footprints and Directional Extension}

The goal of this Section is to give some clarity by formally defining how a mathematician can tell the community what the paper is for and how it is built.
Such statements are informal when made in prose, and the three definitions that follow (Definitions~\ref{def:cat}--\ref{def:de}) are what turns each of them into a statement that is true or false of a formal object.

\begin{definition}[Proof category $\Proof(\Sigma)$]\label{def:cat}
Objects are formulas of $L$. A morphism $A\to B$ is a $\approx$-class of derivations of $B$ from the single hypothesis $A$ using axioms in $\Sigma$. Composition is substitution (cut), and identities are trivial derivations. Several parallel premises are packaged as one conjunction (up to the bureaucratic conjunction packaging of Definition~\ref{def:approx}); no general categorical product property ($\beta/\eta$ for $\wedge$) is assumed. The empty hypothesis is the formula $\top$, so a theorem $T$ is a morphism $\top\to T$.
\end{definition}

\begin{example}\label{ex:cat}
Write $S_{\mathrm{alt}}$ for ``the foot $H$ of the altitude lies strictly between $B$ and $C$, so $d+e=c$'' and $S_{\mathrm{sim}}$ for ``the triangles $ABC$, $HBA$, $HAC$ are similar''. In a typical derivation of the Pythagorean theorem by similar triangles, the two are independent. A chain of three lemmas is
\[
\top \xrightarrow{\,p_1\,} S_{\mathrm{alt}}\wedge S_{\mathrm{sim}} \xrightarrow{\,p_2\,} S_{\mathrm{alt}}\wedge(a^2=cd)\wedge(b^2=ce) \xrightarrow{\,p_3\,} T_N,
\]
where $p_1$ packages two parallel sub-arguments as one segment, $p_2$ is the passage from similarity to proportion, and $p_3$ is the final summation $a^2+b^2=c(d+e)=c^2$. The theorem is the composite $p_3\circ p_2\circ p_1:\top\to T_N$.
\end{example}

\begin{definition}[Footprints]\label{def:fp}
For a derivation $D$ in trimmed form $D^{\mathrm{trim}}$, the \emph{presented footprint} $\Ax(D)\subseteq\Sigma$ is the set of axioms occurring as leaves of $D^{\mathrm{trim}}$. Since trimming and the administrative laws of Definition~\ref{def:approx} preserve the set of axiom leaves reaching the conclusion, $\Ax(p)$ is well-defined on morphisms $p$ in $\Proof(\Sigma)$. For a statement $S$, the \emph{minimal footprints} $\Min(S)$ is the family of $\subseteq$-minimal $\Sigma'\subseteq\Sigma$ with $\Sigma'\vdash S$. This family need not be a singleton, and we never select an element of it.
\end{definition}

Minimal axiom sets are not unique, so no canonical ``minimal hypotheses of $T$'' is used. We use the footprint of the derivation as presented, which is canonical, and the family $\Min(S)$ as a benchmark: a derivation $D$ of $S$ is \emph{footprint-minimal} if $\Ax(D)\in\Min(S)$, and it \emph{can be improved} if some $D'$ has $\Ax(D')\subsetneq\Ax(D)$. Since cut elimination may discard vacuous leaves, $\Ax(\Norm(D))\subseteq\Ax(D)$ in general and the inclusion can be strict. The footprint is that of the derivation as written, which is what a reader reusing it sees.

\begin{definition}[Splitting and Directional Extension]\label{def:de}
Let $p:S_0\to T$ be a morphism of $\Proof(\Sigma)$, with $S_0=\top$ when $T$ is proved from the axioms alone and $S_0=H$ when it is proved under a hypothesis $H$. A \emph{splitting} $\sigma$ of $p$ is a factorization
\[
S_0 \xrightarrow{\,p_1\,} S_1 \xrightarrow{\,p_2\,} \cdots \xrightarrow{\,p_n\,} S_n = T,
\qquad p = p_n\circ\cdots\circ p_1,
\]
in which each $S_i$ is a formula occurrence (or conjunction of parallel formula occurrences) of a trimmed representative derivation of $p$, i.e.\ a cut point, such that $S_{i-1}$ occurs non-vacuously as a hypothesis leaf in $p_i$. For each segment $p_i$, let $L_{p_i}$ be its vocabulary and let
\[
\Dir(p_i)=\{\, I \;:\; I \text{ is an interpretation of } (L_{p_i}, \Ax(p_i))\ \text{in some context } (L',\Sigma')\in\mathcal{U}\,\},
\]
where $\mathcal{U}$ is a fixed universe of target theories. The \emph{Directional Extension} of $p$ along $\sigma$ is
\[
\DE(p,\sigma)\;=\;\bigl(\sigma,\ (\Ax(p_i))_{i\le n},\ (\Dir(p_i))_{i\le n}\bigr).
\]
\end{definition}

The transport functor $I_*:\Proof(L_{p_i}, \Ax(p_i))\to\Proof(L', \Sigma')$ has a well-defined source, and its functoriality is automatic: $I_*(p_j\circ\cdots\circ p_i)=I_*(p_j)\circ\cdots\circ I_*(p_i)$ for any single interpretation $I$ of $\bigl(\bigcup_{k=i}^{j} L_{p_k},\, \bigcup_{k=i}^{j}\Ax(p_k)\bigr)$, because $I_*$ is a functor.

\begin{proposition}\label{prop:de}
(i) (Composition of footprints.) For composable $p:A\to B$ and $q:B\to C$ in $\Proof(\Sigma)$ where the hypothesis $B$ occurs non-vacuously in the trimmed representative of $q$, $\Ax(q\circ p)=\Ax(p)\cup\Ax(q)$; after cut elimination only $\Ax(\Norm(q\circ p))\subseteq\Ax(p)\cup\Ax(q)$ holds. (ii) (DE is not determined by DLI.) A single $D$ admits several splittings, and their families of footprints differ in general.
\end{proposition}

\begin{proof}
(i) Composition is substitution at the non-vacuous hypothesis leaves $B$ of $q$, so every leaf of $p^{\mathrm{trim}}$ and of $q^{\mathrm{trim}}$ reaches $C$ in $(q\circ p)^{\mathrm{trim}}$, giving $\Ax(q\circ p)=\Ax(p)\cup\Ax(q)$; normalization may discard vacuous sub-derivations, leaving an inclusion. (ii) If $p=p_3\circ p_2\circ p_1$ with $\Ax(p_1)=\{\alpha\}$, $\Ax(p_2)=\{\beta\}$, and $\Ax(p_3)=\emptyset$, then the splitting $\sigma_1=(p_1,p_2,p_3)$ has footprints $(\{\alpha\},\{\beta\},\emptyset)$, the splitting $\sigma_2=(p_1, p_3\circ p_2)$ has $(\{\alpha\},\{\beta\})$, and the splitting $\sigma_3=(p_2\circ p_1, p_3)$ has $(\{\alpha,\beta\},\emptyset)$. These families of footprints are distinct while $\DLI(p)$ is the same.
\end{proof}

\noindent\textbf{Plain-English meaning.}
Proposition~\ref{prop:de} captures what happens when authors organize their ``scientific chain of thought'' into published lemmas, exposing the \emph{weak links} (to be improved) and \emph{strong links} (to be reused). Part~(i) shows that whenever two steps $p$ and $q$ are merged into a single step $q\circ p$, their axiomatic hypotheses accumulate ($\Ax(p)\cup\Ax(q)$); moreover, if the written argument relies on an assumption that could have been avoided (so that it vanishes in $\Norm(q\circ p)$), the presented footprint $\Ax$ still makes that dependency visible as a weak link. Part~(ii) shows that even when the conceptual route (the DLI) is completely fixed, choosing how to slice the derivation into lemmas (the Directional Extension) changes what the community can do with it. In the proof's example, exposing the three segments $(p_1, p_2, p_3)$ separately reveals that $p_1$ only requires axiom $\alpha$, $p_2$ only requires $\beta$, and $p_3$ requires no axioms from $\Sigma$ at all. If $\beta$ is a restrictive hypothesis, isolating it in $p_2$ flags $p_2$ as the \emph{weak link} where the community should focus efforts to generalize the theorem, while exhibiting $p_1$ and $p_3$ as \emph{strong links} that can be transported to other settings where $\beta$ does not hold. Bundling $p_1$ and $p_2$ into a single lemma $p_2\circ p_1$ merges their footprints into $\{\alpha,\beta\}$, hiding both where the restrictive assumption is actually used and which sub-argument could have been reused independently.

\subsection{Reading weak and strong links}

The formalism makes the two kinds of links announced above visible. A segment $p_i$ whose footprint $\Ax(p_i)$ is larger than necessary (a derivation that \emph{can be improved} in the sense of Definition~\ref{def:fp}) signals a \emph{potential weak link}: assumptions might be weakened or the conclusion strengthened there. A segment with a small footprint and a rich $\Dir(p_i)$ signals a \emph{potential strong link}: its lemma can be transported into different theories or mathematical domains. Which splitting is published, and which links are declared weak or strong, remains the author's choice and is therefore exposed to the community's criticism (Proposition~\ref{prop:de}(ii)).
This is the formal counterpart of the two criticisms recalled in Section~\ref{sec:history}: a Lean script settles correctness, but neither declares a DLI nor exposes a DE.

\section{The five roles of AI in a research process}
\label{sec:roles}

I examine five roles AI can play in a research process, using the vocabulary of Section~\ref{sec:formal}. Each of them changes who produces which part of the derivation, but none removes the need to declare the Derivation-Level Intension and the Directional Extension of a work.

\paragraph{Potential transformative effects of these tools.}
The open question with the use of LLM or agentic systems is how much it will change the familiarity a community has with a component of a chain of thought. Likewise AlphaZero influenced human go players that are now accepting new openings as having chances to be won, whereas before such openings were considered to be loosing patterns \citep{Zhou2018}.

\subsection{Bilbiographical research}

By nature, AI systems can search and summarize. In the same way Google Search, followed by Google Scholar, short-cut the human walk on the shelves of libraries, these tools will be used to look for existing papers of interest for a given research.
Similarly to automated searches, they are subject to biases \citep{Akbar2022}. There have always been strategies to improve the visibility of a paper, they will have to adapt.

\subsection{Computations and solving: preservation of DLI and DE}
\label{sec:preserve:DLI}

A second use of such tool is, once the chain of thought has been established, to get the resolution (the proof) of some of its steps or components.
This usage brings us back to the discussion about proofs in Lean. Nevertheless, for components of a chain of thought, one can guess that, in general, these Lean scripts will be short.
Moreover, \emph{a clearly motivated scientific chain of thought, whose steps are all Lean-proved, remains understandable}.

Indeed, \emph{this use of AI systems should not change the Derivation-Level Intension nor the Directional Extension of the paper}.
Two LLM-generated or Lean versions of the proof will be in such a case $\approx$-equivalent; the difference is administrative reordering.
It shows the role of the two concepts, and especially of the \emph{Directional Extension} for this kind of use of AI: the domain, the intension, the structure of the proof and the identification of weak and strong segments stays the same.

\subsection{Exploration of chains of reasoning: exploration of Directional Extensions}

A third usage is the exploration of possible scientific chains of thought adapted to the intension.
In our formal words it is programming an AI system to \emph{explore different Directional Extensions}.

This use cannot be avoided but it is more questionable: the fixed cost for humans to put in place a reasonable chain of thought (or realistic DE) increases the probability that the intension has some consistence.
The natural way to make a proof is to imagine the domain, the Derivation-Level Intension to be accurate, needed to solve it, and then to deploy different Directional Extensions, identifying dynamically the weak and strong segments, trying to aling as many string segments as possible.
Sometimes the strategy, i.e. the Directional Extension has to be modified. It is in general when they are too many weak segments.
Some other times the only horizon is to change the DLI.

Being abel to deploy plausible DE at will is part of the causes that make us say that we enter in a era of abundance of proofs.
But before that, \cite[Section 6]{kontorovich2025shape} insists on this point: getting \emph{plausible proofs} easily makes more difficult for readers to quickly judge the value of a paper. 
Before this easiness: the fixed costs of DE (the structure of the proof, and especially identified weak and string segments), offered a guarantee on the upper level: the DLI.
This will probably be less the case when AI systems have been in the loop.

In other words: if the structure of the proof comes from a human long work, then the frame of the proof have good chances to be driven by an intense thinking.
On the opposite, in a world where DE is free, I could assert a theorem and a DLI, and ask to an AI system to give me a DE that is true.
There is good chances it is made of weak segments only: I will have obtained the weakest possible proof of this theorem, and nobody will know it.

The reader can answer that interacting with an AI tool may allow me to recover strong links, curating the proof only I obtain stringer links.
This is correct, but it is indeed the use case of Section \ref{sec:preserve:DLI}.

As a conclusion \emph{the community and journals have to find a way to scrutinise Derivation-Level Intension of papers very carefully}.
Editors should demand a clear description of the DE and its weal and string links, reviewers should focus on being sure weak links cannot be easily improved.

\subsection{Critical judgment}

The fourth usage is radically new: it is to get a judgment from a LLM over a reasoning. Appendix \ref{sec:prompt} gives an example.
This is clearly an innovation: without such tools, judgment during the elaboration of a research was obtained in working groups of academic laboratories, from co-authors or trusted peers.

There is nevertheless \emph{a risk of uniformization} of the production but the popularisation of such an usage of AI-systems, but to some extend, journals' role is very similar, the editing process is operating a standardization of the way papers published in a specific journal are delivering their message to the readers.

A tranformation of this role of journals may be implemented in AI systems (to start with: in prompts) that wil convey:
first the description of the DLI and DE, then a mapping of the DLI on the domains of the Journal, and finally a check of commonalities between segments of the DLI with segments of existing work in the community.

\subsection{Full production is a dream}

The last usage would be a full production of a paper, from the intension to the details of the components, including the scientific chain of thought.
It never happened up to today, since, as previously detailed in this paper, the claimed successes of AI-based provers borrowed the intension (like ``solving Navier-Stokes'') to the collective of mathematicians.
It is difficult to imagine \emph{how a system could generate intension, without a prior human-triggered intension?}
A human input could be: ``draft three possible models or theories that could answer to this specific problem?''.
To shortcut this step, one should design a system that is systematically identifying problems in an ecosystem,\footnote{like ``risk intermediation by a set of investment banks'', or to go away from the theme of the second part of this paper ``electricity production in the area of a datacenter''.} and then launch inputs like the former ones.
The intension will nonetheless belong to the humans who decided to design and run this system targeting this specific ecosystem.

\subsection{What remains human, that is useful}

\emph{The process of research is made of juggling between the feasibility of components of a scientific chain of thought (DE), and the design of this very chain of thought, under the premises of an intension (DLI)}.
This is what allows science to progress thanks to a continuous positive criticism of researchers' efforts.
I have shown that the availability of LLMs and AI Systems is not changing this process, 
and \emph{a paper to be publish should particularly emphasis two points: the intension and the scientific chain of though.
It should make available all elements enabling to check carefully the components of the chain of though.}

\section{Concluding remarks}

I claim that AI is a tool, and that it changes neither the purpose of mathematicians nor their usefulness to society.
It certainly changes the time spent on technicalities, but such shifts are not new: the field has absorbed printing, symbolic computation and numerical experiments without changing what it is for.

A result has always been the outcome of a dialogue between what the authors wish to produce and what they know to be possible.
This dialogue is a closed loop: the statement is adjusted until it can be proved, and the proof in turn reshapes the statement.
This will not change, even if more people become able to close such loops with the help of AI.

What does change is the cost of a proof.
When proofs are cheap, they become abundant \cite{Tao2026Math20}, and if we only attach importance to whether each of them is true or false, this abundance is a huge cloud of dust.
Mathematicians as a community do more than certify truth.
They choose which problems deserve to be stated, relative to a field, and they identify and discuss the weaknesses and strengths of the path chosen to solve them.

This is the only way to extend the realm of what mathematics can deal with towards the ``outside''.
If mathematics is the language of the possibles \cite{bouveresse1988pays}, then attaching what is possible to important questions continues to be the way to foster scientific innovation, since all sciences use this language.
These questions may lie inside the domain of mathematics or at its frontier, but solving them extends that frontier.

To capture what papers have always conveyed, and should continue to convey, I proposed two concepts and formalised them with some rigor:
the \emph{Derivation-Level Intension} (in short, the \emph{Intension}), which is the choice of the problem and the angle from which it is exposed,
and the \emph{Directional Extension}, which in common language is a well-structured chain of reasoning exposing its weak and strong links.
Whoever, or whatever, produces the derivations, a paper that declares these two elements remains open to the criticism of its community, and that criticism is how research is validated and how it moves forward.

\paragraph{Disclosure.}
Large Language Models have been used for reformulation of some paragraphs, to improve their readability or their style.
Most of the formalisation of Section \ref{sec:formal} have been done with the support of different LLMs; the ``announcement subsection \ref{sec:announce}'' have been enterily written by my self and different back and forth with LLMs (of the Directional Extension) have been attempted with different tool until I was happy with it.
I then produced Lean version of Proposition \ref{prop:norm} and \ref{prop:de} using a combination of LLMs that I then included in a noweb code available at \href{https://github.com/ca-lehalle/dlide}{\tt github:DLIDE}.

\bibliographystyle{plainnat} 
\bibliography{intension_paper}   

\appendix
\section{Example of critical judgement prompt}
\label{sec:prompt}

This is given as an illustration:
\begin{aiprompt}\tiny
  Act as a rigorous, highly skeptical mathematical reviewer for a top-tier peer-reviewed journal. Evaluate thep rovided theorem, definitions, and proof with absolute logical precision. 

Structure your critique according to the following 5 analytical phases:

\begin{enumerate}
\item THEOREM STATEMENT \& NOVELTY
  \begin{itemize}
  \item Clarity \& Precision: Is the theorem statement precise, unambiguous, and mathematically sound? Identify any redundant conditions or undefined notation.
  \item  Canonical Context: Is this result already known, or is it a minor variation of an established theorem? List well-known equivalent or closely related standard theorems (e.g., by name or standard textbook result).
  \end{itemize}

\item LOGICAL RIGOR \& PROOF MECHANICS
  \begin{itemize}
  \item Chain of Reasoning: Map out the core logical progression step-by-step.
  \item Flaws \& Overclaims: Identify any logical gaps, circular reasoning, unjustified transitions, or mathematical errors.
  \item Stress Testing / Counterexamples: Attempt to construct edge cases or counterexamples (e.g., boundary conditions, zero/infinity, non-standard domains) where a specific step or claim in the proof fails.
  \item Step-Level Novelty: Are the underlying proof techniques standard textbook tricks, or do they introduce novel mechanisms?
  \end{itemize}

\item LITERATURE \& PRIOR ART
  \begin{itemize}
  \item Novelty of Approach: How novel is the proof strategy compared to established methods in the field?
  \item Prior Art Grounding: Identify specific existing results, standard papers, or foundational theorems that this proof relies on or mirrors. (Provide precise naming of standard theorems or known results rather than generic descriptions).
  \end{itemize}

\item ASSUMPTION \& DEFINITION AUDIT
  \begin{itemize}
  \item Assumption Mapping: Trace each stated assumption to the exact line/step in the proof where it is used.
  \item Redundant vs. Missing Assumptions: Are any explicit assumptions never actually used (superfluous)? Are there implicit assumptions required for the steps to hold that were omitted from the theorem statement (e.g., continuity, measurability, compactness)?
  \item Rigor of Definitions: Are all newly introduced concepts rigorously defined before being used?
  \end{itemize}

\item EXPOSITION \& READABILITY
  \begin{itemize}
  \item Hand-Waving vs. Over-Complication: Flag places where critical algebraic or logical steps are omitted ("left as an exercise to the reader") or where simple steps are unnecessarily obscured by heavy notation/tables/recaps.
  \item Structural Flow: Assess whether non-standard notation, visual aids, or excessive formatting distract from or improve readability.
  \end{itemize}

\end{enumerate}
Finally, render a definitive verdict: [ACCEPT / MAJOR REVISION / REJECT] with a concise summary of the primary reason for your decision.
\end{aiprompt}

\end{document}